\documentclass[9pt,twocolumn,twoside]{osajnl}

\journal{ao} 

\setboolean{shortarticle}{false}

\title{
Mode matching error signals using radio-frequency beam shape modulation}

\author[1,2,*]{A. A. Ciobanu}
\author[1,2]{D. D. Brown}
\author[1,2]{P. J. Veitch}
\author[1,2]{D. J. Ottaway}

\affil[1]{Department of Physics, School of Physical Sciences and The Institute of Photonics and Advanced Sensing (IPAS), The University of Adelaide, SA, 5005, Australia}
\affil[2]{Australian Research Council Centre of Excellence for Gravitational Wave Discovery (OzGrav)}
\affil[*]{Corresponding author: alexei.ciobanu@adelaide.edu.au}

\usepackage{ amssymb }
\usepackage{ amsmath }
\usepackage{ mathtools } 
\usepackage{ graphicx }
\usepackage{ xcolor }
\usepackage{ stfloats } 
\usepackage{ float }
\usepackage{ lipsum }
\usepackage{ tabularx }
\usepackage{ multicol }
\usepackage{ hyperref }
\usepackage{ caption }
\usepackage{ subcaption }
\usepackage{ import } 
\usepackage{ letltxmacro } 
\usepackage{ parskip }
 
\definecolor{red}{RGB}{255,0,0}
\definecolor{green}{RGB}{0,128,0}
\definecolor{orange}{RGB}{255,128,0}
\definecolor{blue}{RGB}{0,128,255}

\def\comments_on{0}
\def\paragraphs_on{0}

\newcommand{\comment}[1]{\if\comments_on1
#1
\fi}

\newcommand{\aaci}[1]{\comment{\textcolor{red}{[AAC] #1}}}

\newcommand{\aaco}[1]{\comment{\textcolor{orange}{[AAC] #1 \mbox{}\\} }}

\newcommand{\aacg}[1]{\comment{\textcolor{green}{[AAC] #1 \mbox{}\\} }}

\newcommand{\ddb}[1]{\comment{\textcolor{red}{[DDB] #1}}}

\newcommand{\tem}[1]{HG$_{#1}$}
\newcommand{\bigO}[1]{\mathcal{O}\!\left(#1\right)}

\newcommand{\p}{\mkern1mu}
\newcommand{\iu}{\mathrm{i}\mkern1mu}
\newcommand{\clf}{400~MHz }
\newcommand{\clfd}{$\delta f_2$ }
\newcommand{\mmp}{\text{M} }
\newcommand{\mma}{\mathcal{M} }

\LetLtxMacro{\oldparagraph}{\paragraph}
\renewcommand{\paragraph}[1]{\if\paragraphs_on1
\oldparagraph{#1}\mbox{}\newline
\fi}

\hypersetup{
    colorlinks,
    linkcolor={red!50!black},
    citecolor={blue!50!black},
    urlcolor={blue!80!black}
}

\begin{abstract}
	
	Precise mode matching is needed to maximize performance in coupled cavity interferometers such as Advanced LIGO.  
	In this paper we present a new mode matching sensing scheme that uses a single radio frequency higher order mode sideband and single element photodiodes. It is first order insensitive to misalignment and can serve as an error signal in a closed loop control system for a set of mode matching actuators. We also discuss how it may be implemented in Advanced LIGO.
	The proposed mode matching error signal has been successfully demonstrated on a tabletop experiment, where the error signal increased the mode matching of a beam to a cavity to 99.9\%.
	
\end{abstract}

\setboolean{displaycopyright}{true}

\begin{document}
	
\maketitle


\section{Introduction}

\begin{figure}[h]
	\centering
	\includegraphics[width=\linewidth]{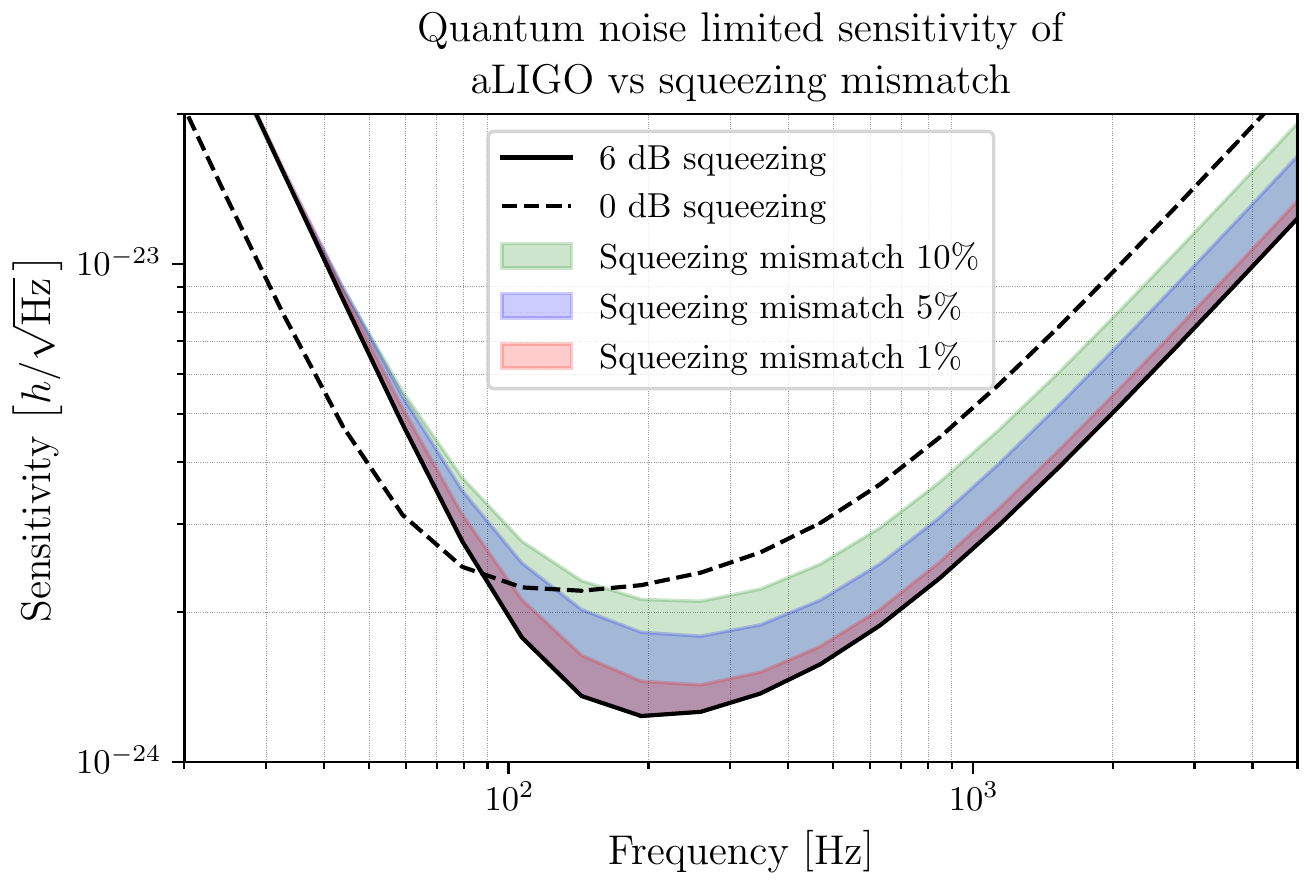}
	\caption{Shown is the reduction in quantum noise limited sensitivity if the squeezer is mismatched by 1\%, 5\%, or 10\% in an aLIGO like detector setup using \textsc{Finesse}~\cite{brown_finesse_2014-1,brown_pykat_2020}. The worst case scenarios are shown as a mismatch between of 10\% can be distributed between beam waist and position mismatch, thus for a given mismatch the sensitivity will lie in the coloured bounds.
	\aaci{What's going on with the y axis unit label? Dan please add script for generating plot into overleaf repo.}
	}
	\label{fig:sqz_vs_loss}
\end{figure}

Precise control of mode matching is desirable in many high precision optical cavity experiments to minimise optical losses. 
This is especially true for advanced gravitational wave (GW) detectors, such as Advanced LIGO~\cite{the_ligo_scientific_collaboration_advanced_2015}, and Advanced Virgo~\cite{the_virgo_collaboration_advanced_2015} that use multiple coupled cavities and squeezed light injection to maximize sensitivity \cite{barsotti_squeezed_2019}.
Fig.~\ref{fig:sqz_vs_loss} depicts the degradation when using 6~dB of squeezing due to varying percentages of mode mismatches.
A mode mismatch of only 10\% will almost completely nullify the expected improvement from squeezing.

\begin{figure}[h]
	\centering
	\includegraphics[width=\linewidth]{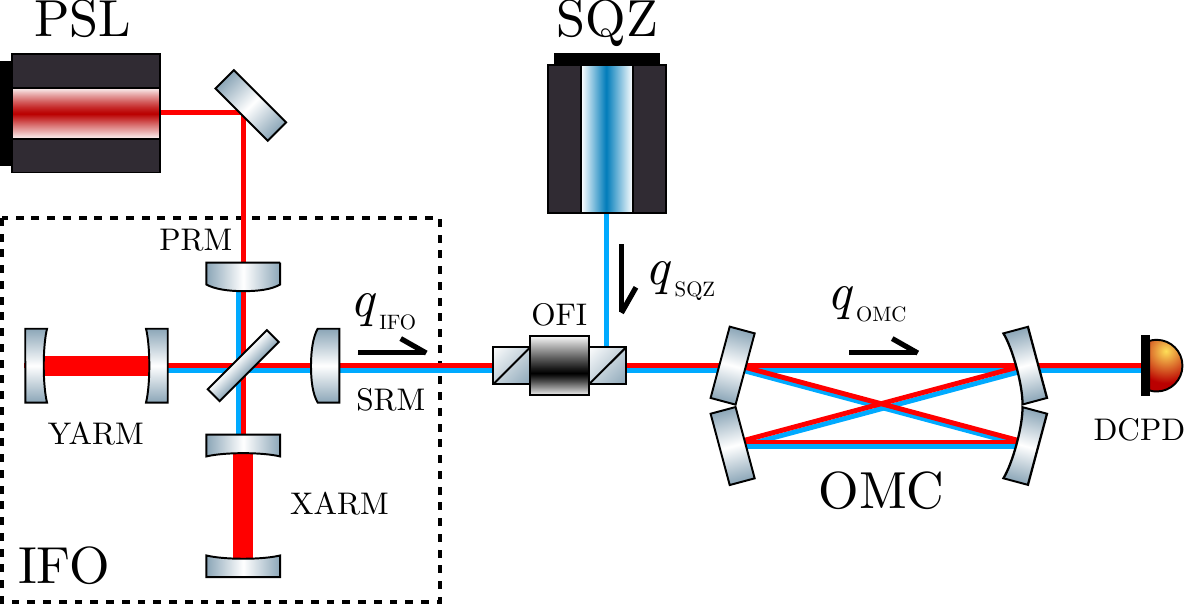}
	\caption{A simplified diagram of a gravitational wave interferometer with squeezing for illustrating the mode matching problem that is being considered in this paper. The interferometer and squeezer beam both need to be simultaneously mode matched to each other and the output mode cleaner (OMC).
	The mode matching problem reduces to making all of the $q$ parameters equal $q_{\text{IFO}} = q_{\text{SQZ}} = q_{\text{OMC}}$.
	The definition of the $q$ parameter and its application to mode matching is given in section \ref{sec:theory}.}
	\label{fig:sqz_to_ifo}
\end{figure}

A diagram of the simplified problem being considered is illustrated in Fig.~\ref{fig:sqz_to_ifo}.
The squeezer here has been simplified to a single laser source (SQZ), which is injected into the main interferometer (IFO) through the output faraday isolator (OFI).
The gravitational wave signal is then filtered by the output mode cleaner (OMC) cavity, whose purpose it is to remove junk light and radio-frequency sidebands that do not contribute to the gravitational readout. 
The GW signal is measured by the photodiode DCPD.

The modes of the OMC cavity, IFO and SQZ beam are parameterized by the $q$ parameters $q_\text{OMC}$, $q_\text{IFO}$, and $q_\text{SQZ}$ respectively.
To minimise the squeezing losses due to mode mismatch it is necessary to minimise the difference between all three $q$ parameters when the beams are overlapped.
Current plans for upgrades and future designs will implement additional optical cavities called \textit{filter cavities}~\cite{mcculler_frequency-dependent_2020}---which also require mode matching to. 
These will provide frequency-dependent squeezing for a broadband reduction in quantum noise, as opposed to frequency-independent squeezing as shown in Fig.~\ref{fig:sqz_to_ifo}. 

Mode matching the squeezer to the IFO becomes increasingly challenging due to the ever increasing requirements for optical power stored in the interferometer. 
The increased power introduces significant thermal lensing which varies over time and must be actively compensated for \cite{brooks_overview_2016}. 
Various actuators are available to correct for this \cite{lawrence_adaptive_2002,rocchi_thermal_2012} however accurate sensors for measuring the distortion to the optical fields are not as prevalent.

Heterodyne wavefront sensing techniques using quadrant-photodiodes can be used to sense both alignment~\cite{mavalvala_experimental_1998} and shape~\cite{magana-sandoval_sensing_2019-1} distortions in optical fields. Annularly segmented diodes, known as \textit{bullseye-detectors}, can also be used~\cite{mueller_determination_2000, nicholas_smith-lefebvre_modematching_2012}. 
However, these schemes require the use of multiple sensors and \textit{gouy-phase telescopes} to sense each of degree of freedom: translation, displacement, size, and curvature of the beam. 
An alternative method proposed by Fulda~et.al.~\cite{fulda_alignment_2017} uses an electro-optic beam deflector to generate a Hermite-Gaussian \tem{10} sideband that can be used as a reference for alignment sensing.
This simplifies an experiment by using single-element photodiodes and does not require any gouy-phase telescopes.
Segmented photodiodes and higher order Hermite-Gaussian modes have additionally been used to generate DC cavity length error signals \cite{shaddock_frequency_1999,miller_length_2014}.

In this paper we demonstrate a new heterodyne scheme that uses a single \tem{20+02} sideband that beats with a \tem{00} carrier to produce mode matching error signals for both degrees of freedom on a single photodiode with no gouy-phase telescopes. 
Experimentally we show how this can be used to match a beam's shape to an optical cavity to 99.9\%. 
This scheme allows multiple sequential cavities to be mode matched together and can be applied for mode matching the squeezed light source, filter-cavity, interferometer, and the OMC in current and future GW detectors.

Firstly, in section~\ref{sec:theory} we present the theory of the mode matching error signal as well as a simpler model in section~\ref{sec:first_order_theory} that can compute the mode matching error signal to first order in mismatch.
In section~\ref{sec:experiment_setup} an experimental demonstration of the proposed mode matching scheme is described followed by an analysis of the results in section \ref{sec:experiment_results}.
Finally, in section~\ref{sec:application_LIGO} we present how this mode matching error signal can be applied to Advanced LIGO~\cite{the_ligo_scientific_collaboration_advanced_2015}.

\section{Theory}
\label{sec:theory}
\subsection{Ideal mode matching error signal}
\label{sec:mismatch_background}
Mode matching, as defined in the context of optical cavities is the process of changing the incoming beam shape to match the eigenmode of a cavity. 
The eigenmode of a stable cavity is parameterized by a single complex number $q = z + \iu z_R$, often referred to as the beam parameter \cite{siegman_lasers_1986}, where $z$ is the distance to the beam waist, and $z_R = \pi w_0^2 / \lambda$ is the Rayleigh range of a beam with waist size $w_0$ and wavelength $\lambda$.

\begin{figure}[t!]
	\centering
	\includegraphics[width=0.7\linewidth]{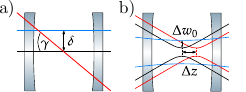}
	\caption{a) The two misalignment degrees of freedom: tilt $\gamma$ (red), and displacement $\delta$ (blue) with respect to the cavity axis (black).
	b) The two mode matching degrees of freedom: waist position $\Delta z$ (red), and waist size $\Delta w_0$ (blue) with respect to the cavity eigenmode (black).}
	\label{fig:mismatch}
\end{figure}

Mode matching is achieved when the waist position and size of the input beam matches the eigenmode of the cavity, which can be done by minimizing the quantities $\Delta w_0$ and $\Delta z$ illustrated in figure~\ref{fig:mismatch}.
These two quantities form the two orthogonal degrees of freedom for mode matching. 
Both must be measured simultaneously in any mode matching sensing scheme.

\paragraph{Ideal mode matching error signal}
An idealised mode matching error signal would indicate for a given $q_1$ and $q_2$ what is the mode mismatch between them, and more importantly the direction in $q$ space along which they lie.
One can define a quantity $\mma$, that has the required properties
\begin{align}
	\mma = \iu \left(\frac{q_2 - q_1 }{q_2 - q_1^*}\right)
	\label{eq:def_ideal_mm_errsig}
\end{align}
where the mode mismatch is given by $\mmp = |\mma|^2$ \cite{bayer-helms_coupling_1984} .
The mode mismatch $\mmp$ between  two Gaussian beam parameters $q_1$ and $q_2$ is defined by the fraction of \tem{00} power in $q_1$ that appears as higher order \tem{} modes in $q_2$.



To first order in mode mismatch $\mma$ is
\begin{align}
	\mma = \frac{\iu \, \varepsilon_q}{2} + \bigO{\varepsilon^2}
\end{align}
where $\varepsilon_q$ is a relative change in beam parameters.
It is defined as
\begin{align}
	\varepsilon_q = \frac{q_2 - q_1}{z_{R,1}} = \varepsilon_z + \iu \varepsilon_{z_R}
	\label{eq:varepsilon_q}
\end{align}
where $z_{R,1}$ is the imaginary part of $q_{1}$\footnote{
The Rayleigh range in \eqref{eq:varepsilon_q} can be for either $q_1$ or $q_2$ as they are approximately equal for small mode mismatches. The lowest numerical error is achieved using $(z_{R,1}+z_{R,2})/2$
}.
The real part of this $\varepsilon_z$ represents a waist position mismatch, and $\varepsilon_{z_{R}}$ waist size mismatch between the $q_1$ and $q_2$; these are also referred to as the orthogonal \textit{mode matching quadratures}. 
We show in section \ref{sec:first_order_theory} that the proposed mode matching error signal approximates the ideal mode matching error signal to first order in mode mismatch up to a constant factor.

\subsection{Single sideband error signal}
\label{sec:first_order_theory}

Here we provide a broad overview of our technique and detailed mathematical derivations can be found in appendix~\ref{sec:full_errsig_derivation}.
In order to understand how our scheme can be applied to a gravitational wave detector we must first cover two key cases.
The first being the mode matching between two beams as in figure~\ref{fig:sqz_to_ifo} between $q_{\text{IFO}}$ and $q_{\text{SQZ}}$.
The second being the mode matching between a beam and a cavity as in figure~\ref{fig:sqz_to_ifo} between $q_{\text{IFO}}$ and $q_{\text{OMC}}$.
Additional instances that require mode matching error signals in advanced LIGO and A+ can be obtained from these two cases and are discussed in more detail in section \ref{sec:application_LIGO}. 
For example mode matching to a filter cavity would require the application of the second case.

For the first case, let us consider the behaviour of a beam consisting of a \tem{00} carrier at the optical frequency $\omega_0$ with a beam parameter $q_1$ and a \tem{02,20} single sideband with a beam parameter $q_2$ at a frequency offset $\Omega$,
\begin{equation}
    E = \left[U_{00}(q_1) + (U_{20}(q_2) + U_{02}(q_2))e^{\iu\Omega t}\right] e^{\iu\omega_0 t}.
    \label{eq:E}
\end{equation}
Here $U_{nm}$ are the 2D transverse mode shape for the $nm$-th Hermite Gaussian modes.
In the case $q_1=q_2$, this beam would look as if the carrier $q_1$ parameter is being modulated in both waist position and size at a frequency $\Omega$. 
This is due to the fact that a \tem{00} beam combined with a small amount of \tem{02} can be considered to be a pure \tem{00} but with a slightly different $q$ parameter~\cite{anderson_alignment_1984} and a single sideband implies a modulation in both quadratures.

\begin{figure}[t]
	\centering
	\includegraphics[width=\linewidth]{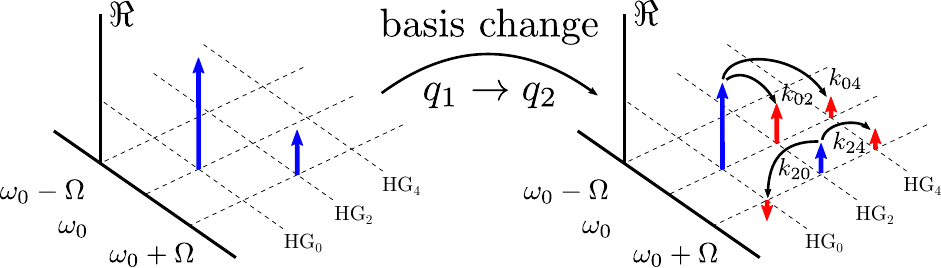}
	\caption{
	An example of a basis change. 
	The two mode spectra on the left and on the right describe the same beam but in bases $q_1$ and $q_2$ respectively. 
	The new Hermite-Gauss modes that are generated during the basis change are indicated with red arrows. 
	The coupling coefficients $k_{nm}$ used to scatter fields to other Hermite-Gaussian modes in a basis change are listed in Table~\ref{tab:coupling_coefficients}}
	\label{fig:basis_change}
\end{figure}

The optical field \eqref{eq:E} incident on a photodiode will generate a photocurrent $\mathcal{I} \propto |E|^2$. To compute the mode matching error signal we must first ensure all optical fields are represented in the same $q$ basis---here we choose to project the \tem{02,20} in the basis $q_2$ into $q_1$, as depicted in figure \ref{fig:basis_change}. 
When the carrier and sideband have the same $q_1=q_2$ parameter the \tem{00} and \tem{02,20} are orthogonal and produce no beat. 
In general when $q_1\neq q_2$ demodulating $\mathcal{I}$ at a frequency $\Omega$ and low-passing the output results in the complex-valued signal
\begin{align}
	\mathcal{Z} &= \sqrt{P_{00}P_{02}} \frac{\sqrt{2}}{4}\varepsilon_{q} +\bigO{\varepsilon^2}.
\end{align}
Here $P_{00}$ and $P_{02}$ are the power in the carrier and reference sideband respectively. The real measured error signal at a demodulation phase $\phi$ is then $\mathrm{Re}(\mathcal{Z}e^{\iu\phi})$. Thus we can easily select either the real or imaginary part of the mode matching~\eqref{eq:varepsilon_q} using the demodulation phase.

\subsubsection{On reflection of a resonant cavity}
\label{sec:second_mm_errsig}

For the second case described in section \ref{sec:first_order_theory} we must consider when \eqref{eq:E} with $q_1 = q_2$ is reflected off of an optical cavity. 
The cavity is held on resonance for the \tem{00} carrier, the resonating field has a shape defined by $q_{cav}$.
The complex-valued error signal is this case is found to be
\begin{align}
	\mathcal{Z} &= \sqrt{P_{00}P_{02}} \left(R_{cav} - 1 \right) \frac{\sqrt{2}}{4}\varepsilon_{q} +\bigO{\varepsilon^2},
\end{align}
where $\varepsilon_{q}$ is the mismatch between $q_1$ and $q_{cav}$, and $R_{cav}$ is the reflectivity of the cavity for the carrier. 
$R_{cav}$ depends on whether the cavity is over-coupled, impedance matched, or under-coupled~\cite{bond_interferometer_2016}
\begin{align}
	R_{cav} = \mathrm{sign}\!\left(R_1 - R_2\right) \left(\frac{R_1 - R_2}{R_1 + R_2 - 2}\right)^2,
	\label{eq:cav_refl_coeff}
\end{align}
where $R_{1,2}$ are the reflectivity of the input and output mirrors. 
For over-coupled, impedance matched, and under-coupled cavities $R_{cav}$ takes the values \mbox{$-1 < R_{cav} < 0$}, $R_{cav}=0$, and \mbox{$0 < R_{cav} < 1$} respectively.

If $q_2 = q_{cav}$ there is again no optical beat measured at the reflected photodiode due to the \tem{} modes of the incident and circulating fields being perfectly orthogonal. 
In the case where the incident beam is mismatched this orthogonality between the modes is broken and an optical beat is measurable. 


\subsubsection{Generating a single \tem{nm} sideband}


There are multiple ways in which a single sideband can be generated. 
In this work we will produce the single sideband field as described by \eqref{eq:E} by using a Mach-Zender interferometer. 
One of the paths will contain a Fabry-Perot cavity which is locked to the \tem{20+02} mode of phase modulated sideband generated by an EOM. 
This is then combined with the original unmodulated carrier field to produce the final field. Although simple the $q$ of the carrier and sideband will not be identical, thus care must be taken to mode match and align them well.

An alternative method would be to use a modulated mode-mismatched beam where the frequency of the sidebands are chosen so that \tem{20,02} higher order mode of sidebands resonates at the same time as the \tem{00} at the carrier frequency. 
Both then have the same $q$ parameter and are transmitted through the cavity.
The downside of this method is that it is not particularly efficient as the vast majority of the power resides in the carrier frequency, the non-transmitted sideband, and in the \tem{00} mode. 
For short cavities the modulations frequencies can also be prohibitively high unless a cavity near to geometric instability is used.

A more efficient approach could be to use an EOM with parabolic electrodes that shapes the electric field inside the electro-optic crystal in such a way that it lenses a beam passing through it---similar to Fulda~et.al's~\cite{fulda_alignment_2017} modulator.
A disadvantage with this method is that the lensing created by such a modulator would likely be almost purely astigmatic, and would require an astigmatic cavity to select one of the modulated axes to be able to sense spherical mode mismatch. \ddb{Alternatively a phase plate could be used}

\begin{figure*}[!ht]
	\centering
	\includegraphics[width=0.7\linewidth]{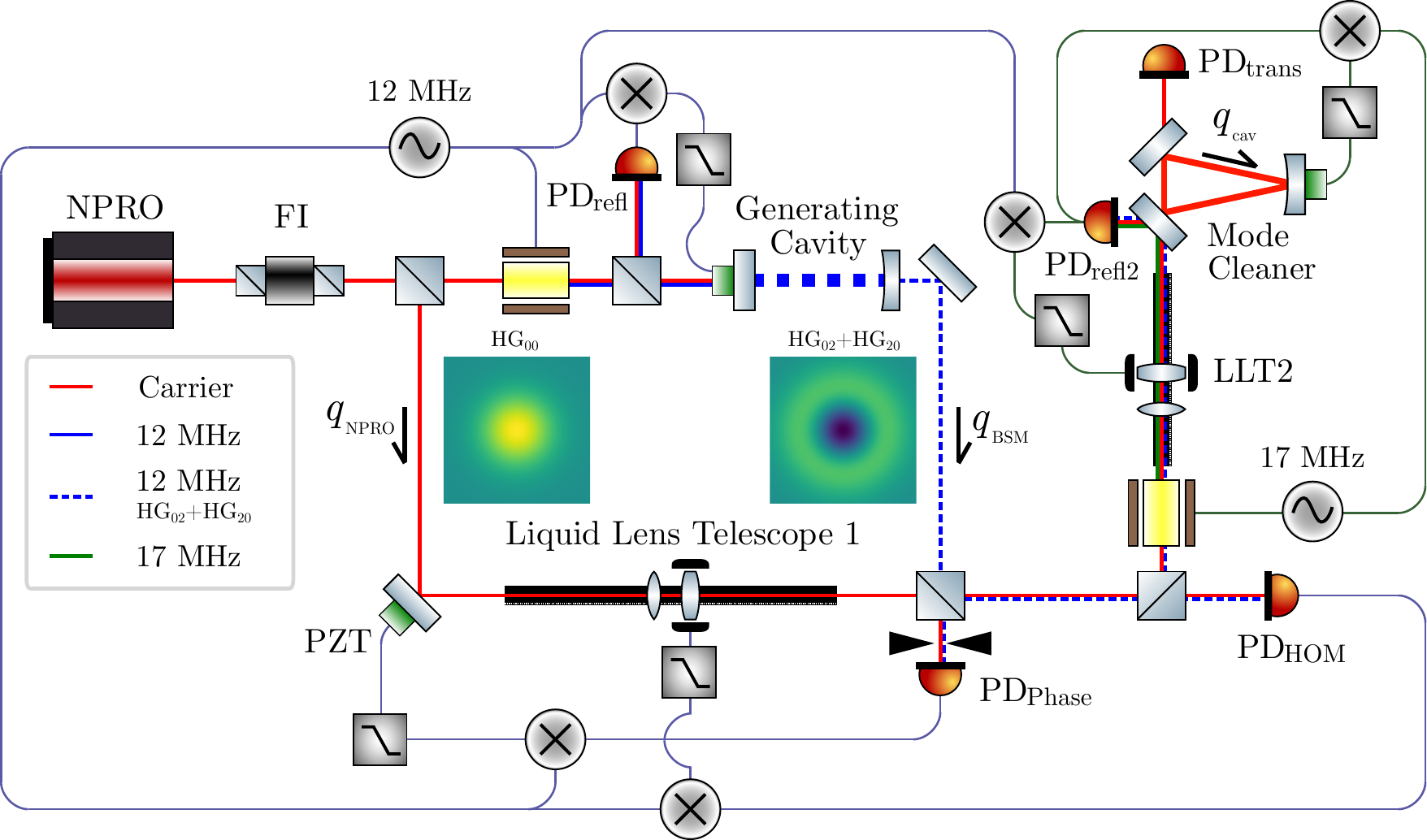}
	\caption{Diagram of the tabletop experiment performed to verify the mode matching error signal. 
	It has been designed to simulate the optical layout of the output of an advanced gravitational wave detector.
	The RF signal generation, mixing, and lowpass filtering was all done on STEMlab 125-14 FPGAs by Red Pitaya running the PyRPL package \cite{neuhaus_pyrpl_2017}.}
	\label{fig:experiment_setup}
\end{figure*}

\section{Experimental Demonstration}
\label{sec:experiment_setup}

To test the validity of the proposed mode matching error signal an experiment as depicted in figure \ref{fig:experiment_setup} has been performed.
It has additionally been designed to replicate the parts of the aLIGO output in figure \ref{fig:sqz_to_ifo} that would be relevant for the mode matching error signal.
The goals of this experiment are to show that zeroing the error signal optimizes the mode matching metrics: maximizing the transmission through a resonant cavity, and minimizing the amount of \tem{02+20} circulating in the cavity.
We show that to achieve this careful consideration of mode matching error signal offsets is required.

A beam from a 1064~nm non-planar ring oscillator (NPRO) is split into two paths in a Mach-Zehnder configuration before being recombined and incident onto a triangular cavity.
One of the paths is intended to represent the interferometer beam, which goes through an adjustable focal length liquid lens (EL-10-30-C by Optotune) on a translation stage.
The NPRO beam is intentionally not spatially filtered in any way in order to mimic the fact the interferometer beam is not a pure \tem{00}~\cite{smith-lefebvre_optimal_2011}.
The liquid lens telescope allows us to change the $q_{\text{NPRO}}$ parameter to match it to $q_{\text{BSM}}$ when the beams combine on the beamsplitter.

The other path goes through the \textit{generating cavity}, which is locked onto the second order transverse mode of one of the 12 MHz phase modulation sidebands that are used to provide a PDH error signal for locking the cavity.
It is worth noting that NPRO beam must be somewhat mode mismatched to this generating cavity to allow sufficient power to build up in the cavity's second order transverse mode.
In this experiment the mode mismatch used for generating the 12 MHz offset second order transverse mode was close to 20\%.


The two arms of the Mach-Zehnder are then recombined with a 50:50 beamsplitter. 
For convenience we will call the beam that is still in the fundamental mode the \textit{optical local oscillator} and the 12 MHz offset second order transverse mode beam as the \textit{mode modulation beam}.
One of the output ports of the Mach-Zehnder is incident onto a photodiode $\text{PD}_\text{phase}$ with an iris. 
The iris clips the beam around the center to break the orthogonality between the optical local oscillator and the mode modulation beam. 
This is used to provide a phase error signal for the microscopic path length changes between the two arms of the Mach-Zehnder.

The phase error signal is used to drive the PZT in the optical local oscillator arm to lock the phases of the two arms. 
This is step is necessary as the phase difference between the optical local oscillator and the mode modulation beam couples into the mode matching error signal readout and hence needs to be kept fixed. 

The photodiode $\text{PD}_\text{HOM}$ measures a mode matching error signal between $q_\text{NPRO}$ and $q_\text{BSM}$ as described in section~\ref{sec:first_order_theory}.
This error signal must be zeroed by adjusting liquid lens telescope~1 in figure \ref{fig:experiment_setup} before the combined beam can accurately measure the mode matching error signal to $q_\text{cav}$ on $\text{PD}_\text{refl2}$.

The combined optical local oscillator and the beam modulation sideband are then incident onto a triangular cavity, which we want to mode match to.
The beam is modulated at 17~MHz to generate the PDH error signal for the mode cleaner, which is measured by demodulating $\text{PD}_\text{refl2}$.
The mode matching error signal for the mode cleaner can then be extracted by locking the cavity to the \tem{00} mode and demodulating $\text{PD}_\text{refl2}$ at 12 MHz as described in section~\ref{sec:second_mm_errsig}.
This signal is then fed back to LLT2 to match to $q_{\text{NPRO}}$ to $q_{\text{cav}}$.
The transmission through the mode cleaner is monitored by the photodiode $\text{PD}_\text{trans}$. A good mode-matching error signal will ensure the transmission is at a maximum when the mode matching error signal is zeroed.

\begin{figure}[!t]
	\centering
	\includegraphics[width=\linewidth]{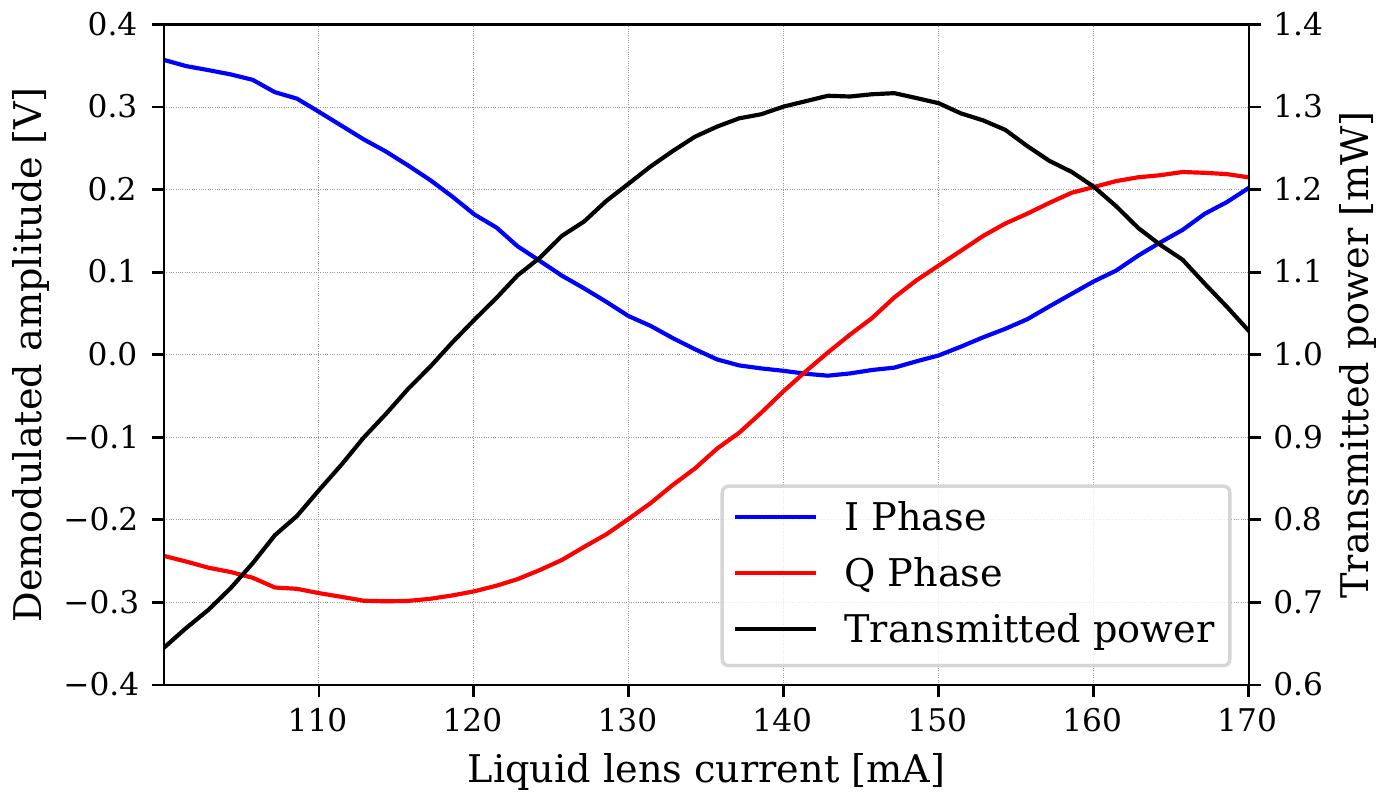}
	\caption{A 1D slice of the mode matching error signal measured at PD$_{\text{refl2}}$ while varying the current on LLT2 in figure \ref{fig:experiment_setup}.}
	\label{fig:1d_errsig_slice_clean}
\end{figure}

\begin{figure*}[!t]
	\centering
	\includegraphics[width=\linewidth]{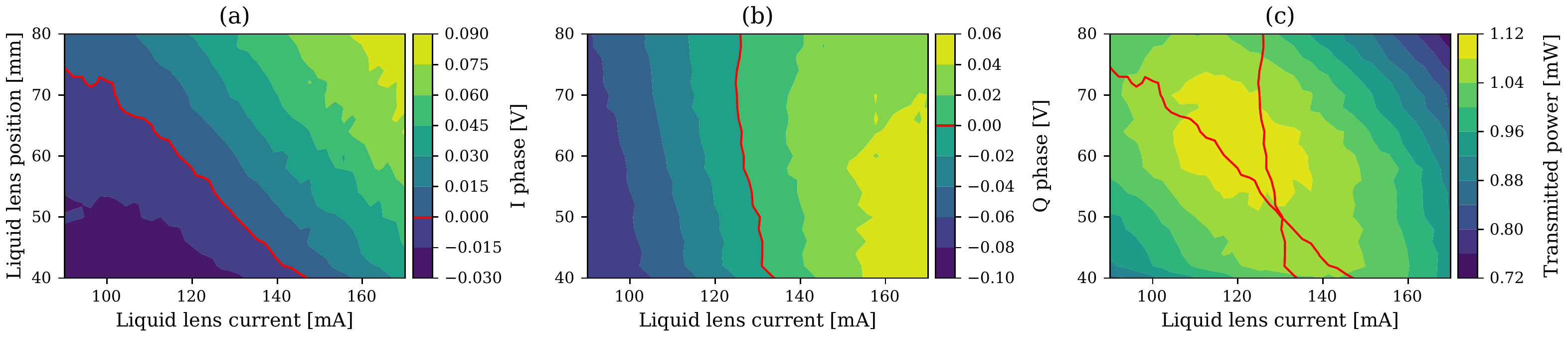}
	\caption{
	a) Measured I phase of the mode matching error signal. The zero crossing is highlighted in red. 
	b) Measured Q phase of the mode matching error signal with the zero crossing highlighted in red. 
	c) The power transmitted through the cavity with both the I and Q crossings overlaid. The offset between the zero crossing and the peak of the transmitted power is a known error signal offset described in section~\ref{sec:offsets}, and results of its subtraction illustrated in figure~\ref{fig:5_point_measurement}.}
	\label{fig:2d_errsig}
\end{figure*}

\section{Experimental Results}
\label{sec:experiment_results}

\paragraph{1D error signal}
\aaco{feels too wordy}
Varying one of the mode matching degrees of freedom provides a one dimensional slice of the mode matching error signal as shown in Fig.~\ref{fig:1d_errsig_slice_clean}. 
The I and Q phase of the mode-matching error signal are each an error signal for a set of two orthogonal mode matching degrees of freedom.
The demodulation phase was chosen such that the Q phase was most sensitive to the liquid lens current.
The I phase shows less response to changing the liquid lens current, especially around the zero of the Q phase error signal. 
The I phase error signal senses a linear combination of liquid lens current and position from the 45 degree contours it makes in the 2D error signal plots in figure \ref{fig:2d_errsig}.a.

\paragraph{2D error signal}
\aaco{Not super happy with this. Not sure why.}
A more complete picture of the mode matching error signal can be obtained by independently scanning both mode matching actuator degrees of freedom in figure \ref{fig:2d_errsig}. 
Here the zero crossings of both I and Q phase form straight lines of different slopes, and hence intersect at a single unique point; the point that the mode matching actuators will be led to.
When overlaid onto the power transmitted from the cavity in figure \ref{fig:2d_errsig}.c it can be seen that the zero of the mode matching error signal is offset from the peak of the transmitted power.

\subsection{Error signal offset}
\label{sec:offsets}
\subsubsection{Generating cavity finesse}
\label{sec:finesse_offset}
The main contributor to the offset shown in figure \ref{fig:2d_errsig}.c is the finite finesse of our generating cavity.
A significant amount of carrier \tem{02} would then leak through the generating cavity on resonance in addition to the 12 MHz \tem{02}.
The carrier and sideband \tem{02} beams produce a beat at 12 MHz which offsets the 12 MHz beat that comes from mode matching error signal.

The offset can be easily measured by demodulating PD$_{\text{refl2}}$ at 12 MHz while blocking the optical local oscillator. 
This beat should read zero if the beam modulation sideband is a pure 12 MHz offset second order mode.
Any other value read is then the offset for the mode matching error signal.

\subsubsection{Misalignment}
\aacg{perfectly fine paragraph}
Misalignment have a quadratic coupling into this mode matching error signal. 
For this experiment it was found that manual alignment by hand produced negligible offsets when compared to the larger sources of offset. 
Due to imperfect alignment through the liquid lens, they induce misalignments whenever their current or position are changed. 
An auto-alignment system could be used to actively compensate for this and reduce misalignment further if required.

\subsubsection{Astigmatism}
\aaco{Kinda hard to follow imo}
It is also possible for astigmatism to present itself as an offset in the mode matching error signal if the cavity being mode matched to is astigmatic compared to the incoming beam or vice versa. 
The error signal will then have a zero between the individual zeroes of the horizontal and vertical planes.
It is possible to extend this mode matching error signal to be able to sense astigmatism by making the cavity that generates the beam modulation sideband to be astigmatic or purely resonate the \tem{20} or \tem{02}.
Then this cavity can be locked to the vertical or horizontal second order modes to provide independent mode matching error signals for either plane.
An astigmatic actuator is then needed to be able to simultaneously drive the error signals in both planes to zero.

\begin{figure*}[!t]
	\centering
	\includegraphics[width=\linewidth]{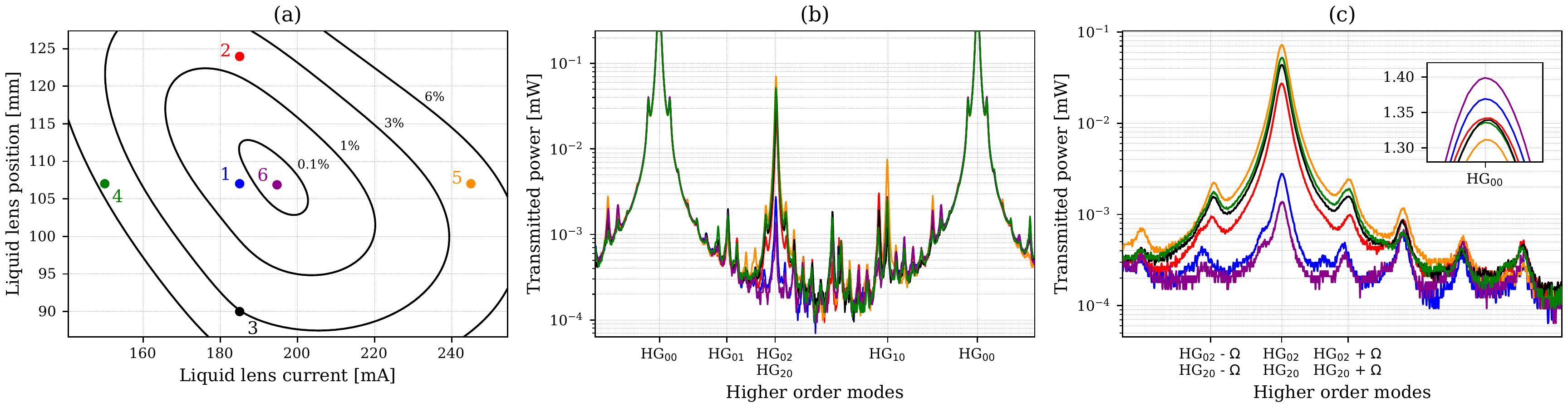}
	\caption{Validating measurement of the mode matching error signal. (a) The six chosen points in actuator space. The contours are the corresponding mode mismatch as predicted by the mode matching error signal. (b) The cavity scans at each point in actuator space. (c) A close up of the \tem{02} resonance of the cavity scans. Inset are the peaks of the \tem{00} resonances. The true mode mismatch is estimated by the ratio of the heights of the \tem{02} and \tem{00} resonances.}
	\label{fig:5_point_measurement}
\end{figure*}

\subsection{Validating measurement}
\aaco{needs to be checked}

\paragraph{cavity scans for validation}
To validate the mode matching error signal it is necessary to compare it to an external measure of mode mismatch such as a cavity length scan by applying a voltage to a cavity mirror PZT.
The ratio of the peak heights of the \tem{02+20} to the \tem{00} resonance provides a measure of mode mismatch.
Performing a cavity scan for each point in actuator space is a time intensive task without an automatic cavity lock acquisition system, as the cavity needs to be relocked to the \tem{00} mode after each scan to read off the mode matching error signal.
It is therefore desirable to obtain a validating measurement of the mode matching error signal with the fewest possible number of points to prevent ambient low frequency fluctuations to influence the results.

\paragraph{5 point measurement setup}
From figure \ref{fig:2d_errsig} we can see that the error signal remains linear over a broad range in actuator space around the perfect mode matching point.
One could then envisage a measurement where a small number of sparsely separated points are measured in actuator space.
Covering the rest of actuator space can be done by fitting a polynomial surface or using an unstructured interpolation method such as thin plate spline interpolation \cite{scipy_10_contributors_scipy_2020}.
We chose the latter due to its simplicity and lack of any external parameters that required optimizing.

With the error signal interpolated across the actuator space the zero of the error signal can be found by brute force evaluation.
It is likely that the interpolated zero point would not zero the error signal, but its measurement can be fed back in to the interpolation to yield an improved estimate of the true zero point of the error signal.

\paragraph{5 point measurement results}
The results of following this procedure are illustrated in figure~\ref{fig:5_point_measurement}.
The first 5 points were chosen from prior assumptions of the location of the zero point of the error signal.
The 6$^{\text{th}}$ point was the zero of the interpolated error signal, which coincided with the true zero of the error signal such that including it in the interpolation did not change the location of the interpolated zero point.

Additional measurements were taken near the zero point (not shown in figure \ref{fig:5_point_measurement}), all yielding higher mode mismatch and non zero error signal values.
This confirms that the offset subtraction in section \ref{sec:finesse_offset} works as intended and the resulting error signal has no offset down to measurement uncertainty. \aaci{what is this uncertainty}
We quote our best mode matching achieved with the mode matching error signal is $99.9^{+0.01}_{-0.1}\%$. \aaci{uncertainty is currently a guess}

We hypothesize that our residual mode mismatch is a result of the difference in astigmatism between the incoming beam and mode cleaner.
Varying the alignment did not affect the height of the \tem{02+20} resonance.\ddb{not sure this paragraph is needed.}

\section{Applying the mode-matching error signal in a gravitational wave detector}
\label{sec:application_LIGO}

\begin{figure*}[!t]
	\centering
	\includegraphics[width=0.7\linewidth]{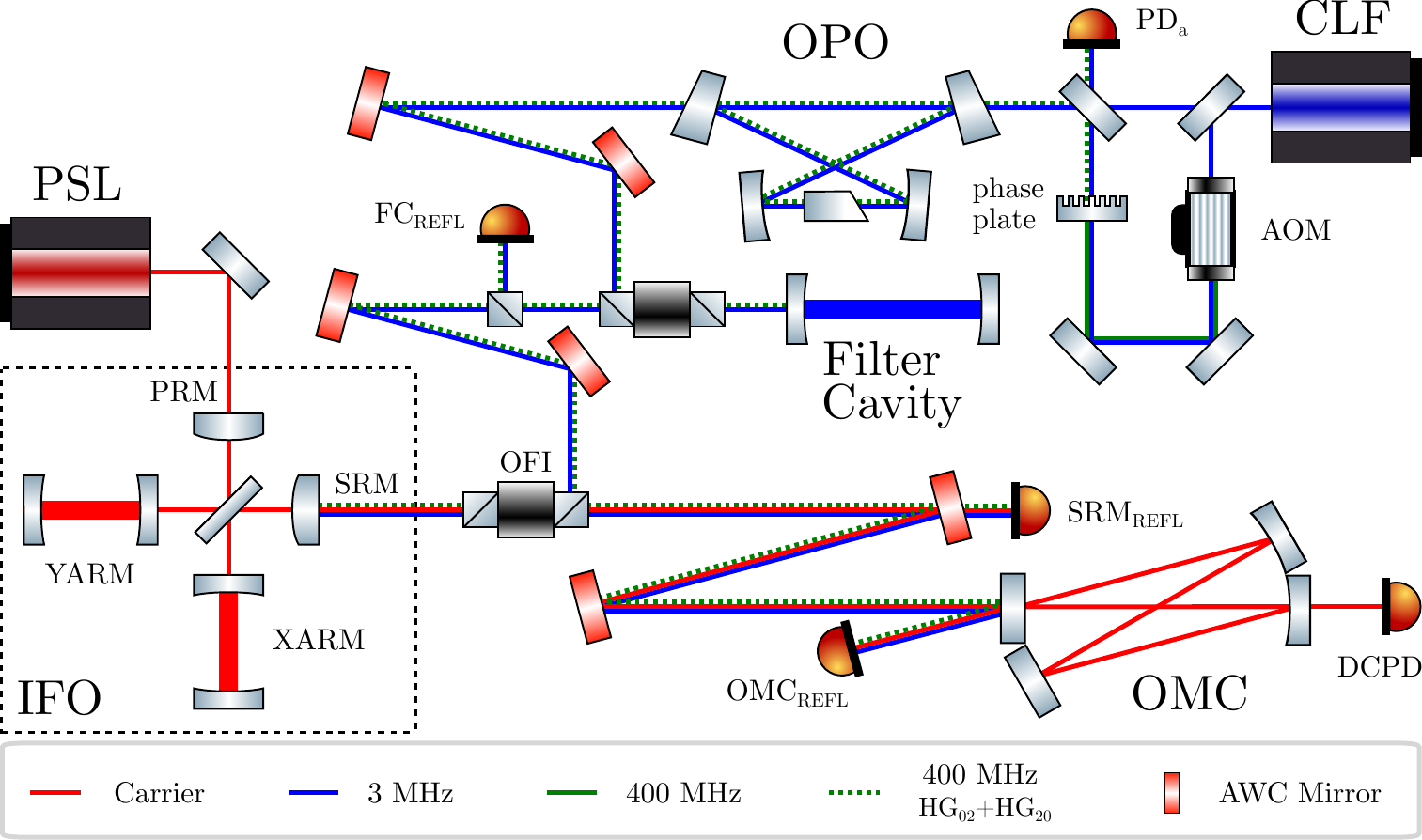}
	\caption{Shown is an optical layout of the output side of the a gravitational wave detector with the key photodiodes and cavities that our scheme would interact with.
	The frequencies listed in the legend are specific to Advanced LIGO.
	}
	\label{fig:aligo_mm_errsig_setup}
\end{figure*}

The proposed mode matching error signal can be applied in a gravitational wave detector to sense the mode mismatch between the cavities in the squeezer path and the interferometer beam, in addition to sensing the mode mismatch between the interferometer beam and the output mode cleaner.
In this section we will consider how this could be implemented in a detector like Advanced LIGO (aLIGO).

A diagram of showing a possible implementation of the proposed mode matching error signal in aLIGO is shown in figure~\ref{fig:aligo_mm_errsig_setup}. The key idea is that the single reference \tem{20+02} sideband can be generated by exciting the higher order mode resonance in the optical parametric oscillator (OPO) that generates the squeezing simultaneously while exciting the \tem{00} that would be the squeezed light mode. As described in section~\ref{sec:second_mm_errsig} this field can then be matched sequentially to the filter cavity, interferometer, and finally the OMC.


To implement the proposed mode matching error signal in aLIGO would require the addition of an AOM to produce a frequency shifted field at \clfd, where \clfd is the frequency offset of the second order transverse modes in the OPO cavity relative to the \tem{00} resonance.
In the absence of a \tem{20,02} phase plate, the CLF beam must be mismatched to enable some coupling into the \tem{20,02}.
The \tem{02, 20} mode of the frequency shifted field would then resonate inside the OPO. 
From an analysis of the OPO cavity geometry \clfd is on the order of \clf in aLIGO.

\subsection{Online mode matching}

To begin considering the mode matching error signal in a gravitational wave detector we start by working from the control light field (CLF) beam in figure \ref{fig:aligo_mm_errsig_setup} towards the OMC.
The CLF is frequency offset locked to the pre-stabilised laser (PSL) and is used for locking the OPO cavity~\cite{mcculler_frequency-dependent_2020}.
It is purposefully frequency offset from the PSL to prevent it from resonating in the interferometer; in LIGO the CLF offset frequency is set at 3~MHz.

The \clf \tem{02} sideband is then incident onto the filter cavity, which is resonant for the 3~MHz CLF \tem{00}.
Demodulating the reflection of the filter cavity on the FC$_{\text{REFL}}$ photodiode at \clf would produce a mode matching error signal between the OPO and the filter cavity as described in section \ref{sec:second_mm_errsig}.
If the filter cavity is not resonant for the CLF but instead some other control field as is in~McCuller~et.al.\cite{mcculler_frequency-dependent_2020}, the mode matching error signal can still be obtained by demodulating at the frequency difference of the filter cavity resonant field and the \clf sideband.

The generated error signal can be used to drive a pair of orthogonal mode matching actuators located between the OPO and filter cavity, labelled as Active Wavefront Control (AWC) mirrors in figure~\ref{fig:aligo_mm_errsig_setup}. 
A number of potential actuator designs have been proposed by the~\cite{cao_high_2020,aidan_brooks_active_2015,kasprzack_performance_2013}.
The proposed error signal is independent of this design and is hence compatible with any of the proposed actuators.

The \clf sideband is combined with the interferometer beam on reflection from the SRM. 
The combined beam will produce a intensity modulation proportional to the mode mismatch between the OPO and the IFO mode as described in section \ref{sec:theory}. This can be demodulated on the photodiode SRM$_{\text{REFL}}$ to be used to drive the pair of AWC mirrors between the filter cavity and the output faraday isolator (OFI). The final stage of the mode matching error signal is to then match the interferometer to the OMC mode using the OMC$_{\text{REFL}}$ photodiode feeding back to the final two AWC actuators.

With all of the previously described mode matching error signals held at their operating point, the entire output chain of the gravitational wave detector should be mode matched \aaci{at what level} to the OMC. The appeal of this mode matching error signal for gravitational wave detectors is its relative simplicity when compared to segmented photodiodes, and a straightforward implementation that requires minimal modification to the existing interferometer.

\subsection{Offline mode matching}

Even without installing the AOM and \clf photodiode electronics an alternative implementation of this mode matching error signal is possible by just locking the OPO cavity to the second order modes and demodulating the same diodes at 3~MHz instead.
The main caveat is that the 3~MHz implementation cannot be run concurrently with squeezing, however this does not pose much of a problem in practice as the squeezer is one of the last subsystems to be turned on in LIGO and the mode matching error signal and optimisations could be run during this initial locking phase.
A bigger issue is that the phase between the CLF and the PSL is locked by demodulating DCPD at 3~MHz \cite{tse_quantum-enhanced_2019,kijbunchoo_low_2020}, which poses a problem as the \tem{02} CLF won't beat with the IFO beam without mode mismatch.
This means that the offline mode matching error signal can only be operated above a certain level of mode mismatch, and could only be used for coarse corrections.

\ddb{On further thoughts, if you run like this you wouldn't be able to use the alignment and phase/frequency offset locking loops, so I don't think this would work actually. Unless you also did some kinf of clipped PD readout somewhere but this would be difficult as it's all in vacuum and little of this light comes out to air.}

\ddb{also requires a purposely mode mismatched CLF}

\subsection{Beacon mode-matching}

For LIGO it is known that the carrier mode differs in shape to the gravitational wave signal mode by some amount due to mode mismatches between the coupled cavities in the Dual-Recycled Fabry-Perot Michelson Interferometer (DRFPMI)~\cite{smith-lefebvre_optimal_2011}.
The proposed mode matching signals do not sense any mode mismatch inside DRFPMI and instead sense mode mismatch between the squeezer and the DC carrier field of the interferometer used for the DC readout~\cite{fricke_dc_2012}.
Therefore it is desirable to achieve as high mode matching as possible for the signal mode and not necessarily the mode that has most of the power.

In order to do this it is necessary to mimic what the gravitational wave does by modulating the arm cavity length to produce a pair of signal sidebands.
The mechanical resonances of the test masses or their suspensions are able to provide a suitable modulation inside the arm cavities that resembles the gravitational wave signal mode.
Mode matching or alignment error signals are then able to lock onto this arm cavity modulation by offsetting the demodulation frequency by the frequency of the arm modulation.
This technique has been used at LIGO under the name of \textit{beacon sensing}~\cite{smith-lefebvre_optimal_2011}, where the arm cavity modulation is being used as an alignment reference instead of the carrier at the anti-symmetric port.
A similar beacon scheme can be employed for the proposed mode matching error signal, which should provide a more accurate mode matching error signal, although at the cost of reduced signal to noise.

It is worth mentioning that some future GW detector designs employ balanced homodyne detection (BHD) for their GW readout, such as A+~\cite{mcculler_design_2018}.
For their implementation of our proposed mode matching error signal beacon sensing provides the only available mode matching reference for the interferometer beam as there is no DC carrier field due to the lack of a DC offset.

\section{Conclusion}
\label{sec:conclusion}

We have proposed and demonstrated a mode matching error signal based on adding a frequency offset \tem{02} single sideband to a main carrier \tem{00} field. 
We have identified and accounted for the main sources of offset in the error signal known to us to as well as how one would go about trying to remove them.
With everything presented here we were able to achieve 99.9\% mode matching of an unfiltered NPRO beam to an optical cavity by following the generated mode matching error signal.
For an aLIGO mode matching error signal it is preferable to be able to mode match the squeezer beam to the gravitational wave detector at LIGO to better than 98\% \cite{barsotti_squeezed_2019}, which seems to be possible using this mode matching error signal given what we have found.

\textbf{\Large Funding.} Australian Research Council (CE170100004);

\textbf{\Large Acknowledgement.} This research was conducted with the support of the Australian Research Council Centre of Excellence for Gravitational Wave Discovery.

\textbf{\Large Disclosures.} The authors declare no conflicts of interest.

\appendix

\section{Gaussian beams}
\aacg{no problems}
The 1D complex amplitude of a Gaussian beam travelling along the $z$-axis with a waist size of $w_0$ located at $z=0$ is given by \cite{siegman_lasers_1986}
\begin{multline}
	u_{n}(x,q) = \left(\frac{2}{\pi}\right)^{1/4} \left(\frac{1}{2^n n! w_0}\right)^{1/2} \left(\frac{\iu z_R}{q}\right)^{1/2} \\ 
	\left(\frac{-q^*}{q}\right)^{n/2} H_n \! \left(\frac{z_R  \, x \sqrt{2}}{w_0 \, |q|} \right) \exp \! \left[-\iu \left(\frac{\pi x^2}{\lambda q}\right)\right]
	\label{eq:u_n_q}
\end{multline}
where $q$ is the complex beam parameter \cite{siegman_lasers_1986}, which encodes all of the necessary information of a beam's shape as
\begin{align}
	q &= z + \iu z_R
	\label{eq:def_q}
\end{align}
where $z$ is the location of the beam waist, and $z_R = \pi w_0^2 / \lambda$ is the Rayleigh range of a beam with waist size $w_0$ and wavelength $\lambda$.

The complex amplitude of a 2~dimensional Hermite-Gaussian \tem{nm} beam with beam shape~$q$ is then given by
\begin{equation}
	U_{nm}(x,y,q) = u_{n}(x,q) \, u_{m}(y,q) .
	\label{eq:u_nm_q}
\end{equation}

To compute the magnitude of coupling of misalignment and mode mismatch to higher order \tem{} modes in section~\ref{sec:first_order_theory} and table \ref{tab:coupling_coefficients} it is convenient to rescale the degrees of freedom to the following relative quantities
\begin{align}
\varepsilon_{\delta} &= \frac{\delta}{w_0}  & \varepsilon_{\gamma} &= \frac{\gamma}{w_0}  \\
\varepsilon_{w_0} &= \frac{\Delta w_0}{w_0}   &  \varepsilon_{z} &= \frac{\Delta z}{z_R}.
\end{align}
For the mode matching degrees of freedom it is also convenient to define a relative change in the $q$ parameter from equation \ref{eq:def_q} as
\begin{align}
	\varepsilon_{q} = \varepsilon_z + \iu \varepsilon_{z_R}
\end{align}
where $\varepsilon_{z_R}$ is the relative change in Rayleigh range, which is given by
\begin{align}
	\varepsilon_{z_R} = \frac{\Delta z_R}{z_{R,1}} = \varepsilon_{w_0}^2.
\end{align}
\section{Mode mismatch}
The mode mismatch between two Gaussian beam parameters $q_1$ and $q_2$ is defined by the fraction of \tem{00} power in $q_1$ that appears as higher order \tem{} modes in $q_2$.
It is given by
\begin{align}
	\mmp = \left|\frac{q_2 - q_1}{q_2 - q_1^*}\right|^2
	\label{eq:def_mode_mismatch}
\end{align}

From equation \ref{eq:def_mode_mismatch} it is evident that it can be satisfied by more than one value of $q_2$ for any $q_1$ and nonzero $\mmp$.
The set of $q_2$'s that satisfy equation \ref{eq:def_mode_mismatch} are called the \mmp \textit{mode matching contour of $q_1$.}
The contour of $q_2$'s for $q_1 = z_1 + \iu z_{R1}$ can be completely described by
\begin{align}
	q_2 =  q_1^* + \frac{2 \, \iu \, z_{R1}}{1 + \iu \sqrt{\mmp} \exp \! \left( \iu \theta \right)}
	\label{eq:def_mm_contour}
\end{align}
where \mbox{$0 \mkern-3mu \le \mkern-3mu \theta \mkern-3mu < \mkern-3mu 2\pi$} is the angular position along the contour of constant mode mismatch \mmp in $q$-space.

\section{Basis change and projections}
\aacg{this is fine}

To describe an incident beam with electric field given by $E(x,y)$ in terms of cavity eigenmodes it is necessary to find the amplitude coefficients $a_{nm}$ of those modes that describe the incident beam by computing the following integral
\begin{equation}
	a_{nm} = \iint_{-\infty}^{\infty} E(x,y) U_{nm}(x,y,q) dy dx
	\label{eq:electric_field_eigenmode_decomposition}
\end{equation}
for all possible combinations of $n$ and $m$. 
However for models with a predominantly \tem{00} field and mismatches less than 10\% it is typically sufficient to compute all the modes with $n+m \le 6$. 

Assuming an ideal \tem{nm} in basis $q_1$ incident on a cavity with eigenmode basis $q_2$ the equation \ref{eq:electric_field_eigenmode_decomposition} can then evaluated by substituting $E(x,y)$ with equation \ref{eq:u_nm_q}.
The resulting integral then becomes the overlap integral of two Hermite Gaussian modes
\begin{equation}
k_{nm} = \int_{-\infty}^{\infty} u_{n}(x,q_1)u_{m}^*(x,q_2) dx
\label{eq:def_scattering_coefficients}
\end{equation}
where if $q_1 = q_2$ this integral reduces to
\begin{equation}
k_{nm} = \delta_{nm}
\end{equation}
where $\delta_{nm}$ is the Kronecker delta. 
For $q_1 \neq q_2$  this integral can be computed either numerically or from an analytic solution \cite{bayer-helms_coupling_1984, brown_finesse_2014}.
In principle this integral needs to be solved for every possible combination of $n$ and $m$, but in practice for small mismatches one only needs to consider the integrals where $|n-m|=2$.
A table of the coupling coefficients to first order in mismatch and misalignment is included in table \ref{tab:coupling_coefficients}.

\begin{table}[!htb]
	\centering
	\caption{\bf First Order Coupling Coefficients}
	\begin{tabular}{p{1.2cm}cc}
		\hline
		Coupling \mbox{Coefficient} & Expression & Notes \\
		\hline
		$k_{n,m,p,q}$ & $k_{n,p} k_{m,q}$ & 2D $\to$ 1D\\
		$k_{n,n}$ & $1+\frac{\iu}{4}\left(2n+1\right) \varepsilon_z +\bigO{\varepsilon^2}$ & $1+$mismatch \\
		$k_{n,n+1}$ & $\left( \varepsilon_\gamma q - \varepsilon_\delta\right)\sqrt{(n+1)} + \bigO{\varepsilon^2}$ & misalignment \\
		$k_{n,n+2}$ & $-\frac{1}{4}\varepsilon_{q}^*\sqrt{\left(n+1\right)\left(n+2\right)} +\bigO{\varepsilon^2} $ & mismatch \\
		$k_{n,m}$ & $-k_{m,n}^* +\bigO{\varepsilon^2} $ & $n \neq m$ \\
		\hline
	\end{tabular}
	\label{tab:coupling_coefficients}
\end{table}

In general the integral in equation \ref{eq:def_scattering_coefficients} appears whenever one wishes to convert the amplitudes of one set of eigenmodes to another set of eigenmodes. 
We refer to this operation as a change of basis, where the basis is parameterised by the complex beam paramter~$q$.

\section{Mode matching error signals}
\label{sec:full_errsig_derivation}
\subsection{Mode matching error signal between two beams}
The proposed mode matching error signal between two beams can be obtained considering a \tem{00} carrier and a beam modulation sideband with powers $P_{00}$ and ${P_{02}}$ respectively.
To compute the interference between the two we choose to project the beam modulation sideband into the carrier basis, which results in an RF \tem{00} component in the carrier basis given by the $k_{0200}$ scattering coefficient from equation \ref{eq:def_scattering_coefficients} to first order in mismatch.

The mode matching error signal $\mathcal{Z}$ is then obtained by demodulating the intensity of the combined beams at the offset frequency, which gives
\begin{align}
	\mathcal{Z} &= \sqrt{P_{00}P_{02}} \left( k_{0200}^* \right) + \bigO{\varepsilon^2}\\
	&= \sqrt{P_{00}P_{02}} \frac{\sqrt{2}}{4}\varepsilon_{q} +\bigO{\varepsilon^2} .
\end{align}

\subsection{Mode matching error signal for resonant cavity}
Consider a beam consisting of a carrier \tem{00} and sideband \tem{02} both in basis $q_1$ incident on a cavity with eigenmode basis $q_{\text{cav}}$.
In order to compute the cavity reflected field it is necessary to project the incident beam into the $q_{\text{cav}}$ basis.
The incident electric field of the incident beam in the $q_{\text{cav}}$ basis is
\begin{align}
	E_{\text{inc}} &= U_{00}(q_{\text{cav}}) \left[\sqrt{P_{00}} k_{0000} + \sqrt{P_{02}} k_{0200} \right] \nonumber \\
	& \qquad + U_{02}(q_{\text{cav}}) \left[\sqrt{P_{00}} k_{0002} + \sqrt{P_{02}} k_{0202} \right] + \bigO{\varepsilon^2}
\end{align}
On reflection all components are promptly reflected with the exception of the resonant mode and frequency, that being the \tem{00} mode at the carrier frequency, which picks up a cavity reflection $R_{\text{cav}}$ defined in equation~\refeq{eq:cav_refl_coeff}.
The reflected electric field is then
\begin{align}
	E_{\text{refl}} &= U_{00}(q_{\text{cav}}) \left[R_{\text{cav}} \sqrt{P_{00}} k_{0000} + \sqrt{P_{02}} k_{0200} \right] \nonumber \\
	& \qquad + U_{02}(q_{\text{cav}}) \left[\sqrt{P_{00}} k_{0002} + \sqrt{P_{02}} k_{0202} \right] + \bigO{\varepsilon^2}
\end{align}
To compute the error signal amplitude we now demodulate at the sideband frequency.
The complex demodulated amplitude is then
\begin{align}
	\mathcal{Z} &= \sqrt{P_{00}P_{02}} \left[R_{\text{cav}} k_{0000}k_{0200}^* + k_{0002}k_{0202}^* \right] + \bigO{\varepsilon^2}\\
	&= \sqrt{P_{00}P_{02}} \left[R_{\text{cav}} k_{0200}^* + k_{0002} \right] + \bigO{\varepsilon^2}\\
	&= \sqrt{P_{00}P_{02}} \left[R_{\text{cav}} k_{0200}^* - k_{0200}^* \right] +\bigO{\varepsilon^2}\\
	&= \sqrt{P_{00}P_{02}} \left( R_{\text{cav}} - 1 \right) \frac{\sqrt{2}}{4}\varepsilon_{q} +\bigO{\varepsilon^2}
\end{align}

\section{Shot Noise}
For a single demodulation of a single optical sideband, the single sided shot noise PSD is given by \cite{harms_quantum-noise_2007}
\begin{align}
	S(f) = 2 \p \hbar \p c \p k \p P
\end{align}
where $P$ is the average incident DC optical power. 
The root mean square noise is then given by
\begin{align}
	\text{RMS}^2 &= \int_{0}^{\Delta f} S(f) df\\
	\text{RMS} &= \sqrt{2 \p \hbar \p c \p k \p P \Delta f}\\
	&= \sqrt{\frac{2 \p \hbar \p c \p k \p P}{T}}
\end{align}
where $\Delta f$ is the integration bandwidth and $T$ is the integration time in seconds.
The signal to noise ratio is then given by
\begin{align}
	\text{SNR} &= \frac{|\mathcal{Z}|}{\text{RMS}} \\
	&= \frac{|\mathcal{Z}|}{\sqrt{2 \p \hbar \p c \p k \p \left(P_{00} + P_{02}\right)}} \sqrt{T}\\
	&= \frac{\sqrt{P_{00}P_{02}}}{\sqrt{{2 \p \hbar \p c \p k \p \left(P_{00} + P_{02}\right)}}} \left|(R_{\text{cav}} - 1) \frac{\sqrt{2}}{4} \varepsilon_q \right| \sqrt{T}
\end{align}
Evaluating the constants and assuming an impedance matched cavity $(R_{\text{cav}} = 0)$ results in
\begin{align}
	&= 5.8\times10^{8} \; \varepsilon_q \left( \frac{\sqrt{P_{00}P_{02}}}{\sqrt{{P_{00} + P_{02}}}} \right) \sqrt{T}
\end{align}
Substituting in the measured powers in the tabletop experiment $P_{00} = 2 mW$, $P_{02} = 20 \mu W$ gives us
\begin{align}
	&= 2.6\times10^6 \; \varepsilon_q \sqrt{T}\\
	&= 3.3\times10^{12} \; \mmp \sqrt{T}
\end{align}
This translates to sub-ppm of mode mismatch even with a 1~ns integration time.
We conclude that in practice technical noise sources and not shot noise will be relevant to the SNR of the mode matching error signal.





\bibliography{my_library}

\begin{thebibliography}{10}
\newcommand{\enquote}[1]{``#1''}

\bibitem{the_ligo_scientific_collaboration_advanced_2015}
{The LIGO Scientific Collaboration}, \enquote{Advanced {LIGO},}
  {\protect\JournalTitle{Classical and Quantum Gravity}} \textbf{32}, 074001
  (2015).

\bibitem{the_virgo_collaboration_advanced_2015}
{The Virgo Collaboration}, \enquote{Advanced {Virgo}: a second-generation
  interferometric gravitational wave detector,}
  {\protect\JournalTitle{Classical and Quantum Gravity}} \textbf{32} (2015).

\bibitem{barsotti_squeezed_2019}
L.~Barsotti, J.~Harms, and R.~Schnabel, \enquote{Squeezed vacuum states of
  light for gravitational wave detectors,} {\protect\JournalTitle{Reports on
  Progress in Physics}} \textbf{82}, 016905 (2019).

\bibitem{mcculler_frequency-dependent_2020}
L.~McCuller, C.~Whittle, D.~Ganapathy, K.~Komori, M.~Tse, A.~Fernandez-Galiana,
  L.~Barsotti, P.~Fritschel, M.~MacInnis, F.~Matichard, K.~Mason, N.~Mavalvala,
  R.~Mittleman, H.~Yu, M.~E. Zucker, and M.~Evans,
  \enquote{Frequency-{Dependent} {Squeezing} for {Advanced} {LIGO},}
  {\protect\JournalTitle{Physical Review Letters}} \textbf{124}, 171102 (2020).

\bibitem{brooks_overview_2016}
A.~F. Brooks, B.~Abbott, M.~A. Arain, G.~Ciani, A.~Cole, G.~Grabeel,
  E.~Gustafson, C.~Guido, M.~Heintze, A.~Heptonstall, M.~Jacobson, W.~Kim,
  E.~King, A.~Lynch, S.~O{\textquoteright}Connor, D.~Ottaway, K.~Mailand,
  G.~Mueller, J.~Munch, V.~Sannibale, Z.~Shao, M.~Smith, P.~Veitch, T.~Vo,
  C.~Vorvick, and P.~Willems, \enquote{Overview of {Advanced} {LIGO} adaptive
  optics,} {\protect\JournalTitle{Applied Optics}} \textbf{55}, 8256 (2016).

\bibitem{lawrence_adaptive_2002}
R.~Lawrence, M.~Zucker, P.~Fritschel, P.~Marfuta, and D.~Shoemaker,
  \enquote{Adaptive thermal compensation of test masses in advanced {LIGO},}
  {\protect\JournalTitle{Classical and Quantum Gravity}} \textbf{19},
  1803--1812 (2002).

\bibitem{rocchi_thermal_2012}
A.~Rocchi, E.~Coccia, V.~Fafone, V.~Malvezzi, Y.~Minenkov, and L.~Sperandio,
  \enquote{Thermal effects and their compensation in {Advanced} {Virgo},}
  {\protect\JournalTitle{Journal of Physics: Conference Series}} \textbf{363},
  012016 (2012).

\bibitem{brown_finesse_2014-1}
D.~D. Brown and A.~Freise, \emph{Finesse} (2014).

\bibitem{brown_pykat_2020}
D.~D. Brown, P.~Jones, S.~Rowlinson, A.~Freise, S.~Leavey, A.~C. Green, and
  D.~Toyra, \emph{Pykat: {Python} package for modelling precision optical
  interferometers} (2020). \_eprint: 2004.06270.

\bibitem{mavalvala_experimental_1998}
N.~Mavalvala, D.~Sigg, and D.~Shoemaker, \enquote{Experimental test of an
  alignment-sensing scheme for a gravitational-wave interferometer,}
  {\protect\JournalTitle{Applied Optics}} \textbf{37}, 7743--7746 (1998).

\bibitem{magana-sandoval_sensing_2019-1}
F.~Maga{\~n}a-Sandoval, T.~Vo, D.~Vander-Hyde, J.~R. Sanders, and S.~W.
  Ballmer, \enquote{Sensing optical cavity mismatch with a mode-converter and
  quadrant photodiode,} {\protect\JournalTitle{Physical Review D}}
  \textbf{100}, 102001 (2019).

\bibitem{mueller_determination_2000}
G.~Mueller, Q.-z. Shu, R.~Adhikari, D.~B. Tanner, D.~Reitze, D.~Sigg,
  N.~Mavalvala, and J.~Camp, \enquote{Determination and optimization of mode
  matching into optical cavities by heterodyne detection,}
  {\protect\JournalTitle{Optics Letters}} \textbf{25}, 266--268 (2000).

\bibitem{nicholas_smith-lefebvre_modematching_2012}
{Nicholas Smith-Lefebvre} and {Negris Mavalvala}, \enquote{Modematching
  feedback control for interferometers with an output mode cleaner,} Tech. Rep.
  P1200034-v3, LIGO (2012).

\bibitem{fulda_alignment_2017}
P.~Fulda, D.~Voss, C.~Mueller, L.~F. Ortega, G.~Ciani, G.~Mueller, and D.~B.
  Tanner, \enquote{Alignment sensing for optical cavities using radio-frequency
  jitter modulation,} {\protect\JournalTitle{Applied Optics}} \textbf{56},
  3879--3888 (2017).

\bibitem{shaddock_frequency_1999}
D.~A. Shaddock, M.~B. Gray, and D.~E. McClelland, \enquote{Frequency locking a
  laser to an optical cavity by use of spatial mode interference,}
  {\protect\JournalTitle{Optics Letters}} \textbf{24}, 1499 (1999).

\bibitem{miller_length_2014}
J.~Miller and M.~Evans, \enquote{Length control of an optical resonator using
  second-order transverse modes,} {\protect\JournalTitle{Optics Letters}}
  \textbf{39}, 2495 (2014).

\bibitem{siegman_lasers_1986}
A.~E. Siegman, \emph{Lasers} (Univ. Science Books, Mill Valley, Calif, 1986).
  OCLC: 14525287.

\bibitem{bayer-helms_coupling_1984}
F.~Bayer-Helms, \enquote{Coupling coefficients of an incident wave and the
  modes of a spherical optical resonator in the case of mismatching and
  misalignment,} {\protect\JournalTitle{Applied Optics}} \textbf{23}, 1369
  (1984).

\bibitem{anderson_alignment_1984}
D.~Z. Anderson, \enquote{Alignment of resonant optical cavities,}
  {\protect\JournalTitle{Applied Optics}} \textbf{23}, 2944--2949 (1984).

\bibitem{bond_interferometer_2016}
C.~Bond, D.~Brown, A.~Freise, and K.~A. Strain, \enquote{Interferometer
  techniques for gravitational-wave detection,} {\protect\JournalTitle{Living
  Reviews in Relativity}} \textbf{19} (2016).

\bibitem{neuhaus_pyrpl_2017}
L.~Neuhaus, R.~Metzdorff, S.~Chua, T.~Jacqmin, T.~Briant, A.~Heidmann, P.-F.
  Cohadon, and S.~Deleglise, \enquote{{PyRPL} ({Python} {Red} {Pitaya}
  {Lockbox}) {\textemdash} {An} open-source software package for
  {FPGA}-controlled quantum optics experiments,} in \emph{2017 {Conference} on
  {Lasers} and {Electro}-{Optics} {Europe} \& {European} {Quantum}
  {Electronics} {Conference} ({CLEO}/{Europe}-{EQEC}),}  (IEEE, Munich,
  Germany, 2017), pp. 1--1.

\bibitem{smith-lefebvre_optimal_2011}
N.~Smith-Lefebvre, S.~Ballmer, M.~Evans, S.~Waldman, K.~Kawabe, V.~Frolov, and
  N.~Mavalvala, \enquote{Optimal alignment sensing of a readout mode cleaner
  cavity,} {\protect\JournalTitle{Optics Letters}} \textbf{36}, 4365 (2011).

\bibitem{scipy_10_contributors_scipy_2020}
{SciPy 1.0 Contributors}, P.~Virtanen, R.~Gommers, T.~E. Oliphant,
  M.~Haberland, T.~Reddy, D.~Cournapeau, E.~Burovski, P.~Peterson,
  W.~Weckesser, J.~Bright, S.~J. van~der Walt, M.~Brett, J.~Wilson, K.~J.
  Millman, N.~Mayorov, A.~R.~J. Nelson, E.~Jones, R.~Kern, E.~Larson, C.~J.
  Carey, {\.I}.~Polat, Y.~Feng, E.~W. Moore, J.~VanderPlas, D.~Laxalde,
  J.~Perktold, R.~Cimrman, I.~Henriksen, E.~A. Quintero, C.~R. Harris, A.~M.
  Archibald, A.~H. Ribeiro, F.~Pedregosa, and P.~van Mulbregt, \enquote{{SciPy}
  1.0: fundamental algorithms for scientific computing in {Python},}
  {\protect\JournalTitle{Nature Methods}} \textbf{17}, 261--272 (2020).

\bibitem{cao_high_2020}
H.~T. Cao, A.~Brooks, S.~W.~S. Ng, D.~Ottaway, A.~Perreca, J.~W. Richardson,
  A.~Chaderjian, and P.~J. Veitch, \enquote{High dynamic range thermally
  actuated bimorph mirror for gravitational wave detectors,}
  {\protect\JournalTitle{Applied Optics}} \textbf{59}, 2784 (2020).

\bibitem{aidan_brooks_active_2015}
{Aidan Brooks}, {Rana Adhikari}, {Stefan Ballmer}, {Lisa Barsotti}, {Paul
  Fulda}, and {Antonio Perreca}, \enquote{Active wavefront control in and
  beyond {Advanced} {LIGO},} Tech. Rep. T1500188-v5, LIGO (2015).

\bibitem{kasprzack_performance_2013}
M.~Kasprzack, B.~Canuel, F.~Cavalier, R.~Day, E.~Genin, J.~Marque, D.~Sentenac,
  and G.~Vajente, \enquote{Performance of a thermally deformable mirror for
  correction of low-order aberrations in laser beams,}
  {\protect\JournalTitle{Applied Optics}} \textbf{52}, 2909 (2013).

\bibitem{tse_quantum-enhanced_2019}
M.~Tse, H.~Yu, N.~Kijbunchoo, A.~Fernandez-Galiana, P.~Dupej, L.~Barsotti,
  C.~D. Blair, D.~D. Brown, S.~E. Dwyer, A.~Effler, M.~Evans, P.~Fritschel,
  V.~V. Frolov, A.~C. Green, G.~L. Mansell, F.~Matichard, N.~Mavalvala, D.~E.
  McClelland, L.~McCuller, T.~McRae, J.~Miller, A.~Mullavey, E.~Oelker, I.~Y.
  Phinney, D.~Sigg, B.~J.~J. Slagmolen, T.~Vo, R.~L. Ward, C.~Whittle,
  R.~Abbott, C.~Adams, R.~X. Adhikari, A.~Ananyeva, S.~Appert, K.~Arai, J.~S.
  Areeda, Y.~Asali, S.~M. Aston, C.~Austin, A.~M. Baer, M.~Ball, S.~W. Ballmer,
  S.~Banagiri, D.~Barker, J.~Bartlett, B.~K. Berger, J.~Betzwieser,
  D.~Bhattacharjee, G.~Billingsley, S.~Biscans, R.~M. Blair, N.~Bode,
  P.~Booker, R.~Bork, A.~Bramley, A.~F. Brooks, A.~Buikema, C.~Cahillane, K.~C.
  Cannon, X.~Chen, A.~A. Ciobanu, F.~Clara, S.~J. Cooper, K.~R. Corley, S.~T.
  Countryman, P.~B. Covas, D.~C. Coyne, L.~E.~H. Datrier, D.~Davis,
  C.~Di~Fronzo, J.~C. Driggers, T.~Etzel, T.~M. Evans, J.~Feicht, P.~Fulda,
  M.~Fyffe, J.~A. Giaime, K.~D. Giardina, P.~Godwin, E.~Goetz, S.~Gras,
  C.~Gray, R.~Gray, A.~Gupta, E.~K. Gustafson, R.~Gustafson, J.~Hanks,
  J.~Hanson, T.~Hardwick, R.~K. Hasskew, M.~C. Heintze, A.~F. Helmling-Cornell,
  N.~A. Holland, J.~D. Jones, S.~Kandhasamy, S.~Karki, M.~Kasprzack, K.~Kawabe,
  P.~J. King, J.~S. Kissel, R.~Kumar, M.~Landry, B.~B. Lane, B.~Lantz,
  M.~Laxen, Y.~K. Lecoeuche, J.~Leviton, J.~Liu, M.~Lormand, A.~P. Lundgren,
  R.~Macas, M.~MacInnis, D.~M. Macleod, S.~M{\'a}rka, Z.~M{\'a}rka, D.~V.
  Martynov, K.~Mason, T.~J. Massinger, R.~McCarthy, S.~McCormick, J.~McIver,
  G.~Mendell, K.~Merfeld, E.~L. Merilh, F.~Meylahn, T.~Mistry, R.~Mittleman,
  G.~Moreno, C.~M. Mow-Lowry, S.~Mozzon, T.~J.~N. Nelson, P.~Nguyen, L.~K.
  Nuttall, J.~Oberling, R.~J. Oram, B.~O{\textquoteright}Reilly, C.~Osthelder,
  D.~J. Ottaway, H.~Overmier, J.~R. Palamos, W.~Parker, E.~Payne, A.~Pele,
  C.~J. Perez, M.~Pirello, H.~Radkins, K.~E. Ramirez, J.~W. Richardson,
  K.~Riles, N.~A. Robertson, J.~G. Rollins, C.~L. Romel, J.~H. Romie, M.~P.
  Ross, K.~Ryan, T.~Sadecki, E.~J. Sanchez, L.~E. Sanchez, T.~R. Saravanan,
  R.~L. Savage, D.~Schaetzl, R.~Schnabel, R.~M.~S. Schofield, E.~Schwartz,
  D.~Sellers, T.~J. Shaffer, J.~R. Smith, S.~Soni, B.~Sorazu, A.~P. Spencer,
  K.~A. Strain, L.~Sun, M.~J. Szczepa{\'n}czyk, M.~Thomas, P.~Thomas, K.~A.
  Thorne, K.~Toland, C.~I. Torrie, G.~Traylor, A.~L. Urban, G.~Vajente,
  G.~Valdes, D.~C. Vander-Hyde, P.~J. Veitch, K.~Venkateswara, G.~Venugopalan,
  A.~D. Viets, C.~Vorvick, M.~Wade, J.~Warner, B.~Weaver, R.~Weiss, B.~Willke,
  C.~C. Wipf, L.~Xiao, H.~Yamamoto, M.~J. Yap, H.~Yu, L.~Zhang, M.~E. Zucker,
  and J.~Zweizig, \enquote{Quantum-{Enhanced} {Advanced} {LIGO} {Detectors} in
  the {Era} of {Gravitational}-{Wave} {Astronomy},}
  {\protect\JournalTitle{Physical Review Letters}} \textbf{123}, 231107 (2019).

\bibitem{kijbunchoo_low_2020}
N.~Kijbunchoo, T.~G. McRae, D.~Sigg, S.~Dwyer, H.~Yu, L.~McCuller, L.~Barsotti,
  C.~Blair, A.~Effler, M.~J. Evans, A.~Fernandez-Galiana, V.~Frolov,
  F.~Matichard, N.~Mavalvala, A.~Mullavey, B.~J.~J. Slagmolen, M.~Tse,
  C.~Whittle, and D.~E. McClelland, \enquote{Low phase noise squeezed vacuum
  for future generation gravitational wave detectors,}
  {\protect\JournalTitle{Classical and Quantum Gravity}}  (2020).

\bibitem{fricke_dc_2012}
T.~T. Fricke, N.~D. Smith-Lefebvre, R.~Abbott, R.~Adhikari, K.~L. Dooley,
  M.~Evans, P.~Fritschel, V.~V. Frolov, K.~Kawabe, J.~S. Kissel, B.~J.~J.
  Slagmolen, and S.~J. Waldman, \enquote{{DC} readout experiment in {Enhanced}
  {LIGO},} {\protect\JournalTitle{Classical and Quantum Gravity}} \textbf{29},
  065005 (2012). ArXiv: 1110.2815.

\bibitem{mcculler_design_2018}
L.~McCuller and L.~Barsotti, \enquote{Design {Requirement} {Document} of the
  {A}+ filter cavity and relay optics for frequency dependent squeezing,} Tech.
  Rep. T1800447, LIGO (2018).

\bibitem{brown_finesse_2014}
D.~D. Brown and A.~Freise, \enquote{Finesse,}  (2014).

\bibitem{harms_quantum-noise_2007}
J.~Harms, P.~Cochrane, and A.~Freise, \enquote{Quantum-noise power spectrum of
  fields with discrete classical components,} {\protect\JournalTitle{Physical
  Review A}} \textbf{76}, 023803 (2007).

\end{thebibliography}


\end{document}